\documentclass[aps,pra,superscriptaddress,twocolumn,amsmath,amssymb]{revtex4-1}
\usepackage{hyperref}
\usepackage{graphicx}
\usepackage{dcolumn}
\usepackage{bm}
\usepackage{hyperref}
\usepackage{color}

\usepackage{multirow}
\usepackage{hhline}

\usepackage[caption=false]{subfig}
\usepackage{braket}
\usepackage{dsfont}

\newcommand\defeq{\mathrel{\overset{\makebox[0pt]{\mbox{\normalfont\tiny\sffamily def}}}{=}}}

\usepackage{microtype}

\newcommand{\BigO}[1]{\ensuremath{\operatorname{O}\left(#1\right)}}

\begin{document}


\title{Realization of arbitrary discrete unitary transformations using spatial and internal modes of light}

\author{Ish Dhand}
\homepage{http://ishdhand.me/}
\email{ishdhand@gmail.com}
\affiliation{
Institute for Quantum Science and Technology, University of Calgary, Alberta T2N~1N4, Canada
}%

\author{Sandeep K.~Goyal}%
\email{sandeep.goyal@ucalgary.ca}
\affiliation{
Institute for Quantum Science and Technology, University of Calgary, Alberta T2N~1N4, Canada
}%

\date{\today}
\begin{abstract}
Any lossless transformation on $n_{s}$~spatial and $n_{p}$~internal modes of light can be described by an $n_{s}n_{p}\times n_{s}n_{p}$~unitary matrix, but there is no known procedure to effect an arbitrary $n_{s}n_{p}\times n_{s}n_{p}$~unitary matrix on light in $n_{s}$~spatial and $n_{p}$~internal modes.
We devise an algorithm to realize an arbitrary discrete unitary transformation on the combined spatial and internal degrees of freedom of light. 
Our realization uses beamsplitters and operations on internal modes to effect arbitrary linear transformations.
The number of beamsplitters required to realize a unitary transformation is reduced as compared to existing realization by a factor $n_{p}^2/2$ at the cost of increasing the number of internal optical elements by a factor of two. 
Our algorithm thus enables the optical implementation of higher dimensional unitary transformations.
\end{abstract}

\pacs{Valid PACS appear here}
\maketitle

\section{Introduction}

Linear optics is important in quantum information processing.
The problem of sampling the output coincidence distribution of a linear optical interferometer, i.e., the BosonSampling problem, is hard to simulate on a classical computer~\cite{Aaronson2013}.
Linear optics enables the efficient simulation of quantum walks~\cite{Lahini2009,Goyal2013,Goyal2015}.
Single-photon detectors and linear optics allow for efficient universal quantum computation~\cite{Knill2001,Kok2007}.

Arbitrary linear optical transformations can be realized on various degrees of freedom (DoFs) of light. 
For instance, any $2\times 2$ unitary transformations on the polarization DoF can be decomposed into elementary operations that are implemented using quarter- and half-wave plates~\cite{Simon1989,Simon1990,Simon2012}.
Any unitary transformation on an arbitrary number of spatial modes can be realized as an arrangement of beamsplitters, phase shifters and mirrors~\cite{Reck1994,Rowe1999,Guise2001} and of temporal modes using nested fiber loops or dispersion~\cite{Motes2014,Motes2015,Pant2015}.
Finally, unitary transformations on orbital-angular-momentum modes of light can be realized using beamsplitters, phase shifters, holograms and extraction gates~\cite{Garcia-Escartin2011}.

Experimental implementations employ spatial modes of light to perform quantum walks~\cite{Peruzzo2010,Crespi2013,Poulios2014}, BosonSampling~\cite{Broome2013,Spring2013,Metcalf2013,Crespi2013a,Bentivegna2015}, bosonic transport simulations~\cite{Harris2015} and photonic quantum gates~\cite{Politi2008,Pooley2012,Meany2015}.
Implementing linear optical transformations on $n$ spatial modes requires aligning $\BigO{n^{2}}$ beamsplitters~\cite{Reck1994}; this requirement poses the key challenge to the scalability of linear optical implementation of unitary transformations.

One approach to overcoming the challenge of realizing a higher number of modes is to use internal DoFs, such as polarization, arrival time and orbital angular momentum, in addition to the spatial DoF. 
In particular, any lossless transformation on $n_{s}$~spatial and $n_{p}$~internal modes can be described by an $n_{s}n_{p}\times n_{s}n_{p}$~unitary transformation.
However, there is no known method to effect an arbitrary $n_{s}n_{p}\times n_{s}n_{p}$~unitary transformation on the state of light in $n_{s}$ spatial and $n_{p}$ internal modes.

Here we aim to devise an efficient realization of an arbitrary unitary transformation using spatial and internal DoFs.
By efficient we mean that the cost of realizing the transformation, as quantified by the number of required spatial and internal optical elements, scales no faster than a polynomial in the dimension of the transformation.
Specifically, we construct an algorithm to decompose an arbitrary $n_{s}n_{p}\times n_{s}n_{p}$ unitary transformation into a sequence of $\BigO{n_{s}^{2}}$ beamsplitters and $\BigO{n_{s}^{2}}$ internal transformations, each of which acts only on the internal modes of light in one spatial mode. 

In contrast to the Reck \emph{et al.}~approach, which allows the realization of any discrete unitary transformation in spatial modes alone, our approach enables the realization into spatial and internal modes
\footnote{The Reck \emph{et al.} procedure decomposes arbitrary $n\times n$ unitary matrices into a product of $2\times 2$ unitary matrices, which are realized as beamsplitters and phase shifters. Hence, the Reck \emph{et al.} procedure cannot incorporate internal degrees of freedom, which require decomposition into $2n_{p}\times 2n_{p}$ beamsplitter matrices and $n_{p}\times n_{p}$ unitary matrices representing internal transformations.
}.
At the cost of increasing the required number of internal optical elements by a factor of two, we reduce the required number of beamsplitters by a factor of $n_{p}^{2}/2$ as compared to the Reck \emph{et al.}~method.
Another difference between our method and the Reck \emph{et al.}~method is that our method requires only balanced beamsplitters, which are easier to construct accurately~\cite{Huisman2014}.

Reducing the required number of beamsplitters at the cost of increasing the number of optical elements is desirable both in free-space and in  on-chip implementations of linear optical transformations. 
Free-space implementations of linear optics require beamsplitters to be stable with respect to each other at sub-wavelength length scales.
On-chip beamsplitters rely on evanescent coupling, which requires overcoming the challenge of aligning different optical channels.
On the other hand, operations on internal elements do not require mutual stability and are typically easier to align and are therefore preferred over beamsplitters.

Moreover, our approach is advantageous experimentally because of its flexibility in the choice of $n_{p}$ and $n_{s}$. 
For instance, consider the realization of a $6\times 6$ unitary matrix.
The Reck \emph{et al.}~approach allows for a realization of this transformation on an interferometer with six spatial modes.
Depending on experimental requirements, our approach allows for a realization of the $6\times 6$ transformations using either (i) six spatial modes ($n_{s} = 6, n_{p}= 1$), (ii) three spatial and two internal modes, for instance polarization ($n_{s} =3, n_{p}= 2$), (iii) two spatial and three internal modes ($n_{s} =2, n_{p}= 3$) or (iv) one spatial and six internal modes ($n_{s} =1, n_{p}= 6$).

Our algorithm is based on the iterative use of the cosine-sine decomposition (CSD).
The relevant background of the CSD is presented in Sec.~\ref{Sec:Background}.
We detail our decomposition algorithm in Sec.~\ref{Sec:Algorithm}.
The cost of realizing an arbitrary unitary matrix is presented in Sec.~\ref{Sec:Cost}.
We conclude with a discussion of our decomposition algorithm in Sec.~\ref{Sec:Conclusion}.

\section{Background: Cosine-Sine Decomposition}
\label{Sec:Background}
In this section, we present the relevant background of the CSD, which is the key building block of our decomposition algorithm.
We describe the factorization of an arbitrary $(m+n)\times (m+n)$ unitary matrix using the CSD.
The section concludes with the realization of a $4\times 4$ unitary transformation on two spatial and two polarization modes of light as enabled by the CSD.

\begin{figure}[h]
 \includegraphics[width=0.49\textwidth]{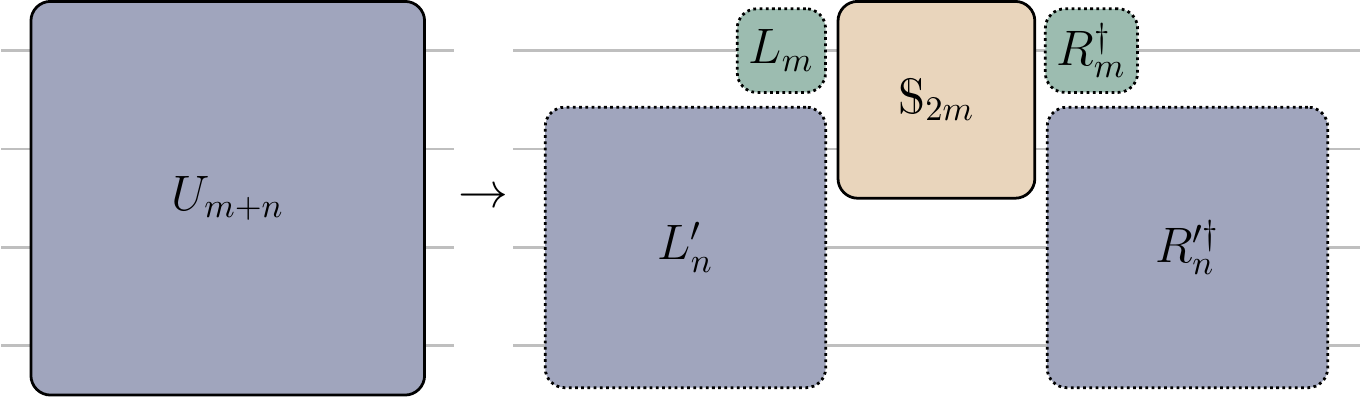}
 \caption{Depiction of the CSD. 
 $U_{m+n}$ is an $(m+n)\times (m+n)$ unitary matrix.
 The CSD factorizes $U_{m+n}$ into the block diagonal matrices represented by $L_{m}, L'_{n},R^{\dagger}_{m},R^{\prime\dagger}_{n}$ and a CS matrix $\mathds{S}_{2m}$~\eqref{Eq:CSMatrix}.
 }
 \label{Fig:CSD}
\end{figure}

The CSD factorizes an arbitrary unitary matrix as follows~\cite{Stewart1977,Stewart1982,Sutton2009}.
For each $(m+n)\times (m+n)$ unitary matrix $U_{m+n}$, there exist unitary matrices $\mathds{L}_{m+n},\mathds{S}_{m+n},\mathds{R}_{m+n}$, such that
\begin{equation}
U_{m+n}= \mathds{L}_{m+n} \left(\mathds{S}_{2m}\oplus \mathds{1}_{n-m}\right)\mathds{R}_{m+n},
\label{Eq:csd}
\end{equation}
where $\mathds{L}_{m+n}$ and $\mathds{R}_{m+n}$ are block-diagonal
\begin{equation}
\mathds{L}_{m+n} = 
\left(\begin{array}{c|c}
L_{m}& {0} \\
\hline
0 & L_{n}'
\end{array}\right),~
\mathds{R}_{m+n} = 
\left(\begin{array}{c|c}
R^{\dagger}_{m}& {0} \\
\hline
0 & R^{\prime\dagger}_{n}
\end{array} \right)
\end{equation}
and $\mathds{S}_{2m}$ is an orthogonal \emph{cosine-sine (CS) matrix} 
\begin{align}
\mathds{S}_{2m} &\equiv \mathds{S}_{2m}(\theta_{1},\dots,\theta_{m})\nonumber\\
&\defeq\left(\arraycolsep=1pt\def\arraystretch{0.9}\begin{array}{ccc|ccc}
\cos\theta_1 & & &\sin\theta_1 & \\
& \ddots & && \ddots &\\
& & \cos\theta_m &&&\sin\theta_m \\
\hline
-\sin\theta_1 & & & \cos\theta_1 & & \\
& \ddots & && \ddots & \\
& & -\sin\theta_m &&&\cos\theta_m
\end{array}\right).
\label{Eq:CSMatrix}
\end{align}
The decomposition of $U_{m+n}$ into $\mathds{L}_{m+n}$, $\mathds{S}_{2m}$ and $\mathds{R}_{m+n}$ is depicted in Fig.~\ref{Fig:CSD}.
Here and henceforth, the respective subscripts of the matrix symbols denote the dimension of the matrix.

The matrices $\mathds{L}_{m+n}$, $\mathds{S}_{2m}$ and $\mathds{R}_{m+n}$ can be constructed using the singular value decomposition as follows.
In order to perform CSD on $U_{m+n}$, we express it as a $2\times 2$ block matrix 
\begin{equation}
 U_{m+n} \equiv\left(\begin{array}{c|c}
  A&B\\\hline
  C&D
 \end{array}\right),
\end{equation}
where $A$ and $D$ are square complex matrices of dimension $m\times m$ and $n\times n$ respectively, and $B$ and $C$ are rectangular with respective dimensions $m\times n$ and $n\times m$.
Each row of the matrix $L_{m}$ ($R_{m}$) is a left-singular (right-singular) vector of $A$, as we prove in Appendix~\ref{Appendix:Construction}.
Similarly, $L^{\prime}_{n}$ and $R_{n}^{\prime}$ are the left- and right-singular vectors of $D$.
Finally, $\{\cos\theta_{i}\}$ is the set of singular values of $A$.
The singular vectors and values of any complex matrix can be computed efficiently using established numerical techniques~\cite{Golub1965,Klema1980,Anderson1992,Press1996}.

\begin{figure}
\subfloat[]{\includegraphics[width=0.49\textwidth]{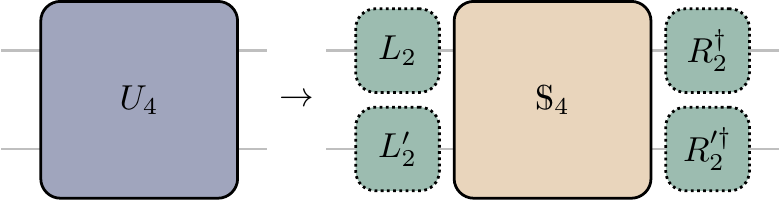}}
 \label{Fig:ns2np2a}\\
\subfloat[]{\includegraphics[width=0.45\textwidth]{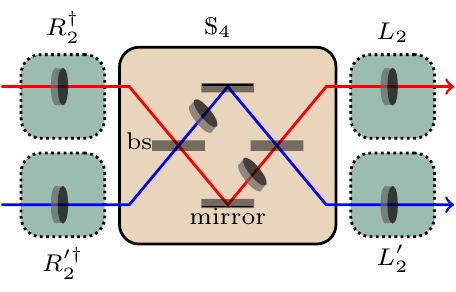}}
  \caption{Realization of a $4\times 4$ unitary matrix $U_{4}$ as a transformation on two spatial and two polarization modes of light. 
 (a) The CSD factorizes $U_{4}$ into the left and right matrices $L_{2}, L'_{2},R^{\dagger}_{2},R^{\prime\dagger}_{2}$ and the CS matrix $\mathds{S}_{4}$~\eqref{Eq:CSMatrix22}.
 (b) The left and right matrices are realized as combinations of quarter- and half-wave plates, and the CS matrix is realized using two balanced beamsplitters and a two wave plates.}
 \label{Fig:ns2np2}
\end{figure}

Now we illustrate the realization of an arbitrary $4\times 4$ unitary matrix as a linear optical transformation on two spatial and two polarization modes~\cite{Goyal2015}. 
The realization is enabled by the CSD, which decomposes the given matrix $U_{4}$ according to 
\begin{equation}
 U_4 = \left(
\begin{array}{c|c}
 L_2 & \\
\hline
& L'_2
\end{array}\right)
\mathds{S}_4\left(
\begin{array}{c|c}
 R_2^\dagger & \\
\hline
& R^{\prime\dagger}_2
\end{array}\right)
\label{Eq:CSD4}
\end{equation}
for $m = n = 2$ as depicted in Fig.~\ref{Fig:ns2np2}(a).
By definition, $U_{4}$ acts on the four-dimensional space $\mathcal{H}_{4}$, which we identify with the combined space
\begin{equation}
\mathcal{H}_{4} = \mathcal{H}^{(s)}_{2}\otimes \mathcal{H}^{(p)}_{2}
\end{equation}
of spatial and polarization modes.
Thus, the $2\times 2$~matrices $L_{2}$ and $R^{\dagger}_{2}$ are identified with transformations acting on the two polarization modes of light in the first spatial mode.
Likewise, $L^{\prime}_{2}$ and $R^{\prime\dagger}_{2}$ correspond to transformations on polarization in the second spatial mode.
Each of these operators $L_{2}, L^{\prime}_{2}, R^{\dagger}_{2},R^{\prime\dagger}_{2}$ can be realized with two quarter-wave plates, one half-wave plate and one phase shifter~\cite{Simon1989,Simon1990}.

 The matrix $\mathds{S}_4$ in Eq.~\eqref{Eq:CSD4} is a CS matrix of the form
\begin{equation}
 \mathds{S}_4(\theta_{1},\theta_{2}) = \left(\begin{array}{cc|cc}
\cos\theta_1 & & ~\sin\theta_1 &\\
&\cos\theta_2 & &~\sin\theta_2\\
\hline
-\sin\theta_1 & & ~\cos\theta_1 & \\
&-\sin\theta_2 & & ~\cos\theta_2
\end{array}\right).
\label{Eq:CSMatrix22}
\end{equation}
This matrix can be decomposed further according to
\begin{equation}
 \mathds{S}_4(\theta_{1},\theta_{2}) = (\mathcal{B}_{2}\otimes \mathds{1}_2)(\Theta_{2}\oplus \Theta_{2}^\dagger)(\mathcal{B}_{2}^\dagger\otimes \mathds{1}_2), 
\label{Eq:SineCosineDecomp}
\end{equation}
where 
\begin{align}
\mathcal{B}_{2} & \defeq\frac{1}{\sqrt{2}}\begin{pmatrix}
1 & i\\
i & 1\\
\end{pmatrix},\label{Eq:BVartheta}\\
 \Theta_{2} &\defeq \begin{pmatrix}
\mathrm{e}^{\mathrm{i}\theta_1} & 0\\
0 & \mathrm{e}^{\mathrm{i}\theta_2}
\end{pmatrix}.
\end{align}
The transformation $(\mathcal{B}_{2}\otimes \mathds{1}_2)$ in Eq.~\eqref{Eq:SineCosineDecomp} represents balanced  beamsplitters, 
whereas, the transformations $\Theta_{2}\oplus \Theta_{2}^\dagger$ can be realized using wave plates acting separately on the polarization of light in the two spatial mode.
Figure~\ref{Fig:ns2np2}(b) depicts the optical circuit for the realization of $U_{4}$ using beamsplitters, phase shifters and wave-plates.

Although the realization of arbitrary $4\times 4$ transformations on two spatial and two polarization modes is known~\cite{Goyal2015}, there is no known realization of an arbitrary $n_{s}n_{p}\times n_{s}n_{p}$ transformation on $n_{s}$ spatial and $n_{p}$ internal modes. 
In the next section, we present a decomposition algorithm to enable this realization. 

\section{Algorithm to design efficient realization}
\label{Sec:Algorithm}

Here we describe the algorithm to decompose an arbitrary unitary matrix into beamsplitter and internal transformations.
Our algorithm is in two parts.
First, we decompose the given unitary matrix into internal transformations and CS matrices. 
Next we factorize the CS matrices into beamsplitter and internal transformations. 
\textsc{matlab} code for the CSD and for our decomposition algorithm is available online~\cite{Dhand2015}.

This section is structured as follows.
Subsection~\ref{Subsec:Inputs} details the inputs and outputs of the decomposition algorithm.
The step-by-step decomposition of the unitary into internal and CS matrices is presented in Subsection~\ref{Subsec:Algorithm}.
The factorization of the CS matrices into elementary operations is described in Subsection~\ref{Subsec:Realization}.

\subsection{Inputs and outputs of algorithm}
\label{Subsec:Inputs}
Here we present the inputs and outputs of our decomposition algorithm.
Our algorithm receives an $n_{s}n_{p}\times n_{s}n_{p}$ unitary matrix as an input.
The algorithm returns a sequence of matrices, each of which describes either a beamsplitter acting on two-spatial modes or an internal unitary operation, which acts on the internal DoF in one spatial modes while leaving the other modes unchanged.
The remainder of this subsection describes the basis and the form of the matrices yielded by our algorithm.

The operators returned by the algorithm act on the combined space
\begin{equation}
\mathcal{H} = \mathcal{H}_{s}\otimes \mathcal{H}_{p},
\end{equation}
where $\mathcal{H}_{s}$ and $\mathcal{H}_{p}$ are spanned 
\begin{align}
\mathcal{H}_{s} &= \operatorname{span}\{\ket{s_{1}},\ket{s_{2}},\dots,\ket{s_{n_{s}}}\}, \\
\mathcal{H}_{p} &=\operatorname{span}\{\ket{p_{1}},\ket{p_{2}},\dots,\ket{p_{n_{p}}}\}
\end{align}
by the $n_{s}$ spatial modes and the $n_{p}$ internal modes respectively for positive integers $n_{s}$ and $n_{p}$.
Each operator acting on the combined state of light can be represented by an $n_{s}n_{p}\times n_{s}n_{p}$ matrix in the combined basis 
\begin{equation*}
\{\ket{c_{k\ell}}\defeq \ket{s_{k}}\otimes\ket{p_{\ell}}: k \in \{1,\dots,n_{s}\},~\ell \in \{1,\dots,n_{p}\}\}
\end{equation*}
of the spatial and the internal modes.
Our algorithm returns the matrix representations of the operators in this combined basis $\{\ket{c_{k\ell}}\}$.

The matrices returned by the algorithm represent either internal or beamsplitter transformations. 
Each internal transformation acts on the internal state of light in a spatial mode but not on the light in the other spatial modes.
In the composite basis, the internal transformations acting on the $k$-th spatial mode are represented as
\begin{equation}
U^{(k)}_{n_{p}} \defeq \mathds{1}_{n_{p}(k-1)}\oplus U_{n_{p}}\oplus\mathds{1}_{n_{p}(n_{s}-k)}
\label{Eq:InternalMatrix}
\end{equation}
for $n_{p}\times n_{p}$ unitary matrix $U_{n_{p}}$.

The algorithm also returns beamsplitter matrices, which mix each of the corresponding internal modes of light in two spatial modes.
The matrix representation of this operator in the composite basis is given by
\begin{equation}
\mathcal{B}^{(k)}_{2n_{p}}\defeq \mathds{1}_{n_{p}(k-1)}\oplus \left(\mathcal{B}_{2}\otimes\mathds{1}_{n_{p}}\right)\oplus
 \mathds{1}_{n_{p}(n_{s}-k-1)}
\label{Eq:BSMatrix}
\end{equation}
for $\mathcal{B}_{2}$ as defined in Eq.~\eqref{Eq:BVartheta} representing a balanced beamsplitter.
To summarize, the algorithm returns a sequence of matrices, each of which is an internal transformation in the form~of Eq.~\eqref{Eq:InternalMatrix} or is a balanced beamsplitter transformation in the form of Eq.~\eqref{Eq:BSMatrix}.

\subsection{Decomposition of unitry matrix into internal and CS matrices}
\label{Subsec:Algorithm}
In this subsection, we present the first stage of our algorithm.
This stage decomposes the given unitary matrix into matrices representing internal transformations~\eqref{Eq:InternalMatrix} and CS transformations
 \begin{equation}
\begin{split}
 \mathds{S}^{(k)}_{2n_{p}}(\theta_{1},\dots,\theta_{n_{p}})\defeq & \mathds{1}_{n_{p}(k-1)}\oplus \mathds{S}_{2n_{p}}(\theta_{1},\dots,\theta_{n_{p}})\\&\oplus
 \mathds{1}_{n_{p}(n_{s}-k-1)},
 \label{Eq:CSMatrixx}
 \end{split}
 \end{equation}
 which enact the CS matrix $\mathds{S}_{2n_{p}}\equiv\mathds{S}_{2n_{p}}(\theta_{1},\dots,\theta_{n_{p}})$~\eqref{Eq:CSMatrix} on the internal degrees of light in two spatial modes without affecting the light in other modes.

\begin{figure}
\subfloat[]{\includegraphics[width=0.49\textwidth]{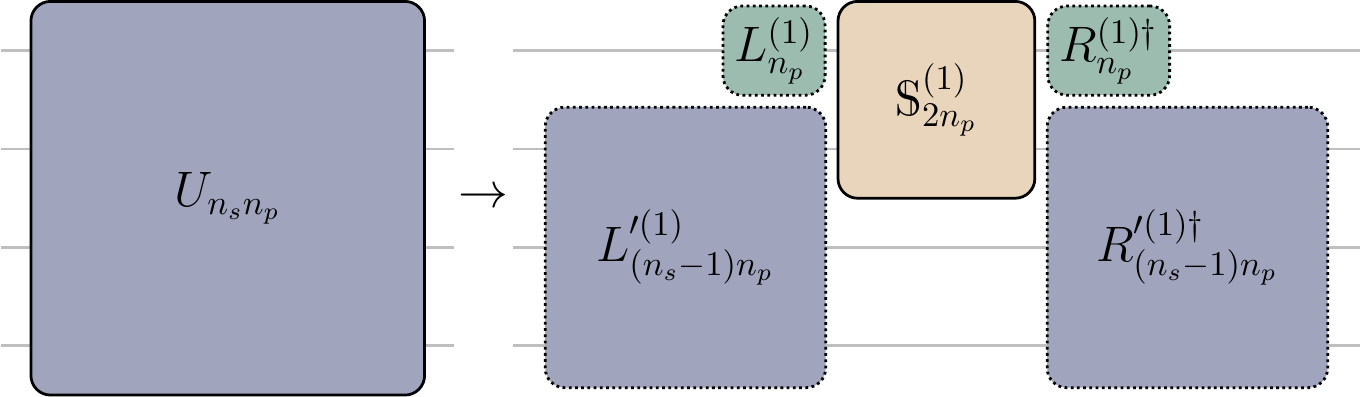}}\\
\subfloat[]{\includegraphics[width=0.49\textwidth]{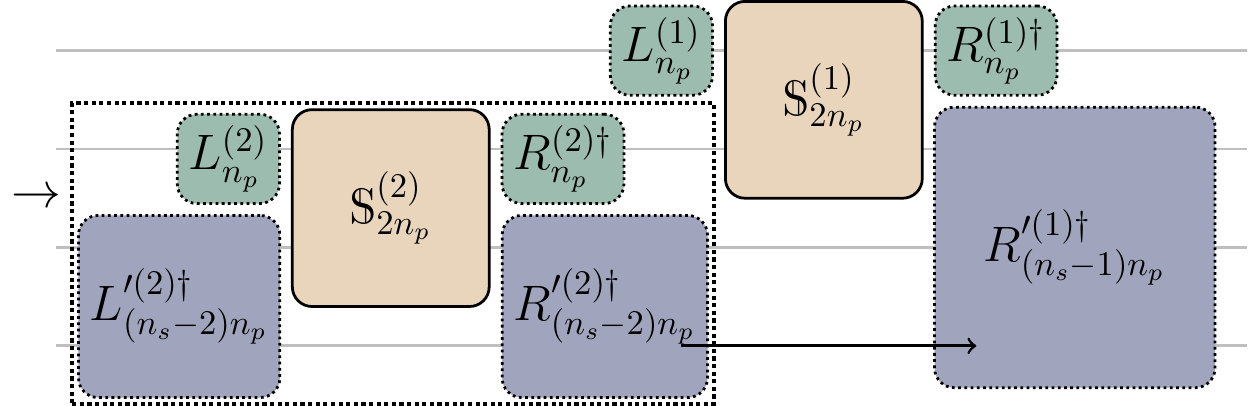}}\\
\subfloat[]{\includegraphics[width=0.49\textwidth]{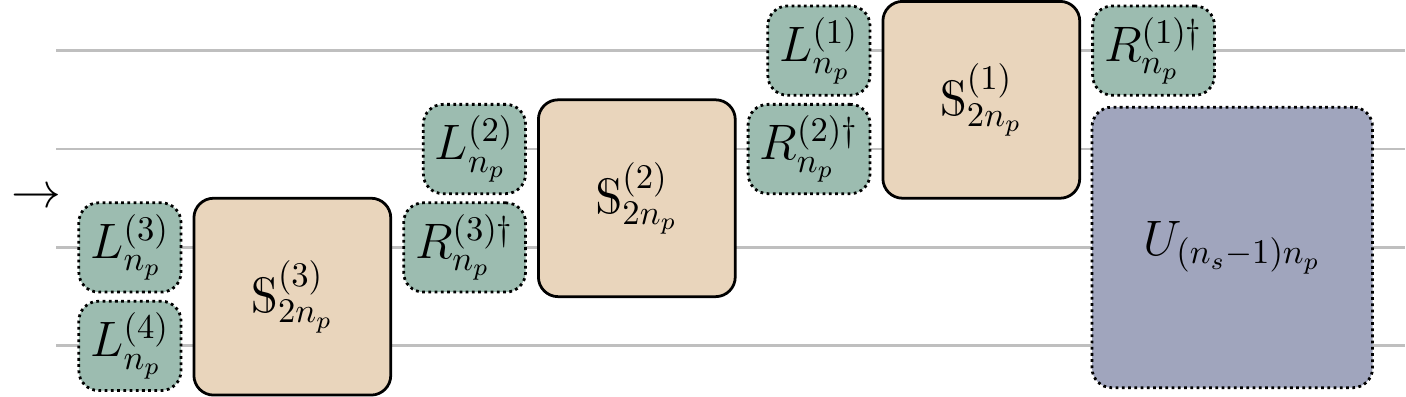}}
\caption{A depiction of the first iteration of our algorithm for the decomposition of a given unitary $U_{n_{s}n_{p}}$ into internal (green) and CS (brown) matrices.
(a) First, the $U_{n_{s}n_{p}}$ unitary matrix is CS decomposed into (i) a $2n_{p}\times 2n_{p}$ CS matrix $\mathds{S}^{(1)}_{2n_{p}}$ acting on the first two spatial modes,
(ii) internal unitary matrices $L_{n_{p}}^{(1)}$ and $R_{n_{p}}^{(1)\dagger}$, each of which act on the internal degrees of the first spatial mode and 
(iii) left and right unitary matrices $L_{(n_{s}-1)n_{p}}^{\prime(1)}$ and $R_{(n_{s}-1)n_{p}}^{\prime(1)\dagger}$ acting on the remaining $n_{s}-1$ spatial modes. 
 (b) The matrix $L_{(n_{s}-1)n_{p}}^{\prime(1)}$ is further CS decomposed. 
 The resultant $R_{(n_{s}-2)n_{p}}^{\prime(2)\dagger}$ from the second decomposition commutes with CS matrix $\mathds{S}^{(1)}_{2n_{p}}$ and can thus be absorbed into $R_{(n_{s}-1)n_{p}}^{\prime(1)\dagger}$. 
(c) The algorithm repeatedly decomposes the left unitary matrices. 
The resultant right unitary matrices are absorbed into the initial right unitary matrix. 
 At the end of one iteration, the algorithm decomposes $U_{n_{s}n_{p}}$ unitary operation into CS matrices, internal unitary matrices and the matrix $U_{(n_{s}-1)n_{p}}$.
 The next iteration of the algorithm decomposes the smaller $U_{(n_{s}-1)n_{p}}$ unitary matrix.
 }
 \label{Fig:FirstStep}
\end{figure}

The first stage comprises $n_{s}-1$ iterations.
Of these, the first iteration factorizes the given $n_{s}n_{p}\times n_{s}n_{p}$ unitary matrix into a sequence of internal and CS matrices and one $(n_{s}-1)n_{p}\times (n_{s}-1)n_{p}$ unitary matrix.
This smaller unitary matrix is factorized in the next iteration.
Figure~\ref{Fig:FirstStep} depicts the first of the $n_{s}-1$ iterations that comprise the first stage.

In general, the $j$-th iteration receives an $(n_{s}+1-j)n_{p}\times (n_{s}+1-j)n_{p}$ unitary matrix.
This iteration decomposes the received unitary matrix into a sequence of internal and CS matrices, and a smaller $(n_{s}-j)n_{p}\times (n_{s}-j)n_{p}$ unitary matrix which is decomposed in the next iteration.

Now we describe the $j$-th iteration of the decomposition algorithm in detail. 
First, the given unitary matrix $U_{(n_{s}+1-j)n_{p}}$ is CS decomposed by setting $m = n_{p}$ and $n = (n_{s}-j)n_{p}$ in the CSD.
This CSD yields the following sequence of matrices 
\begin{align}
U_{ (n_{s}+1-j)n_{p}}=&\, \mathds{L}_{n_{p} + (n_{s}-j)n_{p}} (\mathds{S}_{2n_{p}}\oplus \mathds{1}_{(n_{s}-1-j)n_{p}})\nonumber\\
&\times\mathds{R}_{n_{p}+ (n_{s}-j)n_{p}},
\label{Eq:csd2}
\end{align}
for block diagonal unitary matrices
\begin{align}
\mathds{L}_{n_{p} + (n_{s}-j)n_{p}} &= 
\left(\begin{array}{c|cc}
L_{n_{p}}& \multicolumn{2}{c}{0} \\
\hline
\multirow{2}{*}{0} & \multicolumn{2}{c}{\multirow{2}{*}{$~L^{\prime}_{(n_{s}-j)n_{p}}$}} \\
& \multicolumn{2}{c}{}            \\
\end{array}\right),\nonumber\\
\mathds{R}_{n_{p} + (n_{s}-j)n_{p}} &= 
\left(\begin{array}{c|cc}
R^{\dagger}_{n_{p}}& \multicolumn{2}{c}{0} \\
\hline
\multirow{2}{*}{0} & \multicolumn{2}{c}{\multirow{2}{*}{$~R^{\prime\dagger}_{(n_{s}-j)n_{p}}$}} \\
& \multicolumn{2}{c}{}\\
\end{array}\right),
\end{align}
and orthogonal CS matrix~$\mathds{S}_{2n_{p}}$.

In other words, the first CSD of the $j$-th iteration factorizes the received unitary transformation acting on $n_{s}+1-j$ spatial modes into 
(i)~a $2n_{p}\times 2n_{p}$ CS matrix $\mathds{S}_{2n_{p}}$ acting on the $j$-th and $(j+1)$-th spatial modes, 
(ii)~internal unitary matrices $L_{n_{p}}$ and $R_{n_{p}}^{\dagger}$, each of which act on the internal degrees of the $j$-th spatial mode and 
(iii)~left and right unitary matrices $L_{(n_{s}-j)n_{p}}^{\prime}$ and $R_{(n_{s}-j)n_{p}}^{\prime\dagger}$ acting on the remaining $n_{s}-j$ spatial modes. 
Figure~\ref{Fig:FirstStep}(a) depicts this first CSD for the first iteration.

Next the matrix $L_{(n_{s}-j)n_{p}}^{\prime}$ is CS decomposed.
The resultant $R_{(n_{s}-j-1)n_{p}}^{\prime\dagger}$ from this second CSD commutes with CS matrix $\mathds{S}_{2n_{p}}$ yielded by the first CSD~\footnote{The transformations $R_{(n_{s}-k-1)n_{p}}^{\prime\dagger}$ and $\mathds{S}_{2n_{p}}$ act on mutually exclusive spatial modes so their action is independent of the order of enacting the transformations.}. 
Hence, the operators $R_{(n_{s}-j-1)n_{p}}^{\prime\dagger}$ and $\mathds{S}_{2n_{p}}$ can be swapped, following which we multiply $R_{(n_{s}-j-1)n_{p}}^{\prime\dagger}$ by $R_{(n_{s}-j)n_{p}}^{\prime\dagger}$.
Figure~\ref{Fig:FirstStep}(b) depicts this second round of CSD and of the multiplication of the two right matrices.

The left unitary matrices thus obtained are repeatedly factorized using the CSD. 
The resultant right unitary matrices are absorbed into the initial right unitary matrix $R_{(n_{s}-1)n_{p}}^{\prime\dagger}$. 
Thus, we are left with internal and CS matrices and with a unitary matrix 
\begin{equation}
U_{(n_{s}-j)n_{p}}= \prod_{\ell = 0}^{n_{s}-j-1} R_{(n_{s}-j-\ell)n_{p}}^{\prime\dagger}
\end{equation} obtained by multiplying each of the right unitary matrices.
This completes a description of the $j$-th iteration of the algorithm.

\begin{figure*}
\includegraphics[width = \textwidth]{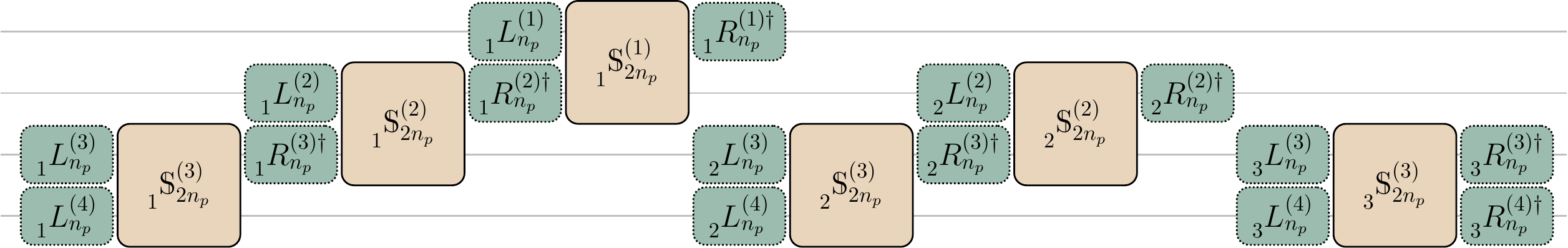}
\caption{A depiction of the output of the first stage of our decomposition algorithm (Subsection~\ref{Subsec:Algorithm}) for the case of $n_{s} = 4$~spatial modes and $n_{p}$~internal modes.
The given $4n_{p}\times 4n_{p}$~unitary matrix is decomposed into $4^{2} = 16$~internal matrices (green) and $n_{s}(n_{s}-1)/2 = 6$~CS matrices (brown). As usual, the right subscript of the matrices is the dimension of the space that the respective operators act on. The right superscript represents the spatial mode that the operators act on. The left subscript specifies the index of iteration that constructed the respective matrices. 
}
\label{Fig:Final}
\end{figure*}

In summary, at the end of the $j$-th iteration, the algorithm decomposes the received $U_{(n_{s}+1-j)n_{p}}$ transformation into internal and CS matrices and $U_{(n_{s}-j)n_{p}}$ as depicted in Fig.~\ref{Fig:FirstStep}(c).
The $(j+1)$-th iteration of the algorithm receives this smaller $U_{(n_{s}-j)n_{p}}$ unitary matrix and decomposes it into internal and CS matrices and an even smaller unitary matrix.
The algorithm iterates over integral values of $j$ ranging from $1$ to $n_{s}-1$.
Figure~\ref{Fig:Final} depicts the output of the algorithm at the end of the final, i.e., $(n_{s}-1)$-th, iteration.
This completes a description of the first stage of the algorithm.

At the end of the first stage, the given unitary matrix has been factorized into a sequence of internal~\eqref{Eq:InternalMatrix} and CS matrices~\eqref{Eq:CSMatrix}.
The internal matrices can be implemented using optical elements if a suitable realization is known for the internal DoF;
such realizations are known for polarization~\cite{Simon1989,Simon1990}, temporal~\cite{Motes2014} and orbital-angular-momentum~\cite{Garcia-Escartin2011} DoFs.
n the next subsection, we present a realization of the CS matrix using beamsplitters acting on spatial modes and internal transformations.

\subsection{Decomposition of CS unitary matrix into elementary operators}
\label{Subsec:Realization}
Here we show how the CS matrices can be decomposed into a sequence of beamsplitter transformations and internal unitary matrices. 
Specifically, we construct a factorization of any $2n_{p}\times 2n_{p}$ CS matrix $\mathds{S}_{2n_{p}}$, which is in the form of Eq.~\eqref{Eq:CSMatrix}, into a sequence of two balanced beamsplitter matrices and two internal-transformation matrices.

Our decomposition of the CS matrix relies on the following identity
\begin{equation}
\mathds{S}_{2n_{p}}(\theta_{1},\dots,\theta_{n_{p}}) = (\mathcal{B}_{2}\otimes \mathds{1}_{n_{p}})(\Theta_{n_{p}}\oplus \Theta_{n_{p}}^\dagger)(\mathcal{B}_{2}^\dagger\otimes \mathds{1}_{n_{p}}), 
\label{Eq:SineCosineDecompNp}
\end{equation}
where $\mathcal{B}_{2}\otimes \mathds{1}_{n_{p}}$ represents a balanced beamsplitter~\eqref{Eq:BVartheta}
and
\begin{equation}
 \Theta_{n_{p}} \defeq \begin{pmatrix}
\mathrm{e}^{\mathrm{i}\theta_1} & &\\
& \ddots &\\
& & \mathrm{e}^{\mathrm{i}\theta_{n_{p}}}
\end{pmatrix}.
\label{Eq:ThetaDefine}
\end{equation}
is a transformation on the internal modes.
Thus, any CS matrix can be realized using two balanced beamsplitters and two internal transformations.

To summarize, the first stage of our algorithm decomposes the given unitary matrix into internal~\eqref{Eq:InternalMatrix} and CS matrices~\eqref{Eq:CSMatrixx}.
The next stage factorizes the CS matrices returned by the first stage into internal and beamsplitter~\eqref{Eq:BSMatrix} transformations, thereby completing our decomposition algorithm.

\section{Cost Analysis: Number of optical elements in realization} 
\label{Sec:Cost}
Here we discuss the cost of realizing an arbitrary $n_{s}n_{p}\times n_{s}n_{p}$ unitary matrix using our decomposition, where the cost is quantified by the number of optical elements required to implement the matrix.
Optical elements required by our decomposition algorithm include balanced beamsplitters, phase shifters and elements acting on internal modes.
We conclude this section with a specific example of decomposing a $2n\times 2n$ transformation into spatial and polarization DoFs.
In this case, our decomposition reduces the required number of beamsplitters to half with the additional requirement of wave plates as compared to using only spatial modes.

Consider the decomposition of an arbitrary $n_{s}n_{p}\times n_{s}n_{p}$ unitary transformation. 
Realization of this transformation using the Reck~\emph{et~al.} method requires $n_{s}n_{p}$ spatial modes and $n_{s}n_{p}(n_{s}n_{p}-1)/2$ biased beamsplitters~\cite{Reck1994}.
In comparison, our decomposition requires $n_{s}(n_{s}-1)$ beamsplitters.
Thus, we reduce the number of beamsplitters required to realize an $n_{s}n_{p}\times n_{s}n_{p}$ transformation by a factor of
\begin{equation}
\eta = \frac{n_{s}n_{p}(n_{s}n_{p}-1)/2}{n_{s}(n_{s}-1)} > n_{p}^{2}/2.
\end{equation}

Although our decomposition reduces the required number of beamsplitters, the number of optical elements required for internal transformations increases by a factor of $2$.
The Reck~\emph{et al.}~approach requires $n_{s}n_{p}(n_{s}n_{p}+1)/2$ phase shifters to effect an $n_{s}n_{p}\times n_{s}n_{p}$ unitary transformation on spatial modes.

Our approach relies on decomposing to beamsplitter and internal unitary transformations, so we count the number of internal optical elements required in our transformation.
Realizing an $n_{p}\times n_{p}$ internal transformation typically requires $n_{p}^{2}$ internal optical elements~\cite{Simon1990,Garcia-Escartin2011,Motes2014}.
Our decomposition requires $n_{s}^{2}$ arbitrary internal transformations, which are represented by matrices $\{L_{n_{p}},L^{\prime}_{n_{p}},R_{n_{p}},R^{\prime}_{n_{p}}\}$ in the output.
These arbitrary transformations can be realized using a total of $n_{s}^{2}n_{p}^{2}$~internal optical elements.
Furthermore, our decomposition also requires $n_{s}(n_{s}-1)$~internal transformations in the form of $\Theta_{n_{p}}$~\eqref{Eq:ThetaDefine}.
Each of these transformations can be realized using $n_{p}$ optical elements for the polarization, temporal and orbital angular momentum modes~
\footnote{%
For the  polarization DoF the $\Theta_{n_{p}=2}$ matrix can be constructed using two elements: a quarter-wave plate and a phase shifter. 
Similarly, for the temporal DoF, the matrix $\Theta_{n_{p}}$ can be realized by setting the reflectivity of the variable beamsplitter to zero and the transmission amplitude to $\mathrm{e}^{\mathrm{i}\theta_{j}}$ at an appropriate time~\cite{Motes2014}. 
The matrix $\Theta_{n_{p}}$ for the orbital-angular-momentum DoF of light can be constructed using a spatial light modulator (hologram)~\cite{Flamm2013}. 
In all these realizations of the matrix $\Theta_{n_{p}}$ no more than $n_{p}$ optical components are required.}
.
In summary, our decomposition requires a total of $n_{s}n_{p}(n_{s}n_{p}+ n_{s}-1)$, which is an increase by a factor
\begin{equation}
\xi = \frac{n_{s}n_{p}(n_{s}n_{p}+ n_{s}-1)}{n_{s}n_{p}(n_{s}n_{p}+1)/2} = 2 + \BigO{1/n_{p}}
\end{equation}
over the cost of the Reck~\emph{et~al.} approach.

Now we consider the example of using polarization as the internal DoF.
Specifically, we compare the cost of realizing an arbitrary $2n\times 2n$ transformation using (i) the Reck~\emph{et~al.} approach on only spatial modes and (ii) our decomposition on the spatial and polarization modes of light, i.e., $n_{s} = n$ and $n_{p} = 2$.
The Reck~\emph{et~al.} decomposition requires $2n$ spatial modes, $n(2n-1)$ beamsplitters and $n(2n+1)$ phase shifters.
In comparison, our approach requires $n(n-1)$ balanced beamsplitters, $n^{2}$ phase shifters and $3n(n-1)/2$ wave plates. 
Thus, our decomposition reduces the required number of beamsplitters and phase shifter by a factor of $2$ each at the expense of an additional $3n(n-1)/2$ wave plates.

To summarize this section, our realization of an arbitrary $n_{s}n_{p}\times n_{s}n_{p}$ unitary matrix reduces the number of beamsplitters required by a factor of at least $n_{p}$.
This completes the analysis of the cost of our decomposition.

\section{Conclusion}
\label{Sec:Conclusion}
In conclusion, we devise a procedure to efficiently realize any given $n_{s}n_{p}\times n_{s}n_{p}$ unitary transformation on $n_{s}$ spatial and $n_{p}$ internal modes of light. 
Our realization uses interferometers composed of beamsplitters and optical devices that act on internal modes to effect the given transformation.
Such interferometers can be characterized by using existing procedures~\cite{Laing2012,Dhand2015a} based on one- and two-photon interference on spatial and internal DoFs~\cite{Walborn2003,Schuck2006,Nagali2009,Karimi2014}.
We thus enable the design and characterization of linear optics on multiple degrees of freedom.

We overcome the problem of decomposing the given unitary transformation into internal transformations by performing the CSD iteratively.
We also open the possibility of using an efficient iterative CSD in problems where the single-shot CSD is currently used~\cite{Bullock2004,Khan2006,Shende2006}.

By employing $n_{p}$ internal modes, the number of beamsplitters required to effect the transformation is reduced by a factor of $n_{p}^{2}/2$ at the cost of increasing the number of internal elements by a factor of $2$.
Our procedure facilitates the realization of higher dimensional unitary transformations for quantum information processing tasks such as linear optical quantum computation, BosonSampling and quantum walks.

\section*{Acknowledgments}
We thank Hubert de Guise, Alexander I. Lvovsky, Barry C. Sanders and Christoph Simon for valuable comments. ID acknowledges AITF, NSERC and USARO for financial support. SKG is supported by NSERC.
\appendix

\section{Construction}
\label{Appendix:Construction}
In this appendix, we present our construction of the CSD. 
Recall that our CSD procedure is a building block of our main decomposition algorithm, which is discussed in Section~\ref{Sec:Algorithm}. 
Although this procedure matches the output of existing procedures~\cite{Stewart1977,Stewart1982}, our procedure emphasizes the key role of the singular value decomposition in the CSD.
Furthermore, numerical implementations of our CSD procedure are expected to be more efficient and stable as compared to existing procedures because of the efficiency and stability of established singular-value-decomposition algorithms~\cite{Golub1965,Klema1980}. 
Note that efficiency of numerical implementations refers to the computational cost of performing the decomposition and differs from the requirement of efficient realization, which deals with the number of optical elements required to experimentally realize the matrices.

First, we recall that the singular value decomposition factorizes any $m\times n$ complex matrix $M$ into the form
\begin{equation}
M = W\Lambda^{M} V^{\dagger}
\end{equation}
for $m\times m$ unitary matrix $W$, $n\times n$ unitary matrix $V$ and real non-negative diagonal matrix $\Lambda^{M}$. 
The matrices $W$ and $V$ diagonalize $M\,M^\dagger$ and $M^\dagger M$ respectively.
In other words, the rows of $W$ and $V$ are the eigenvectors of $M\,M^{\dagger}$ and $M^{\dagger}M$.
These rows are called the left- and right-singular vectors of $M$. 

Now we describe the CSD of a given $(m+n) \times (m+n)$ unitary matrix $U$.
In order to perform CSD of this matrix, we express it as a $2\times 2$ block matrix
\begin{equation}
 \label{Eq:U-Block}
 U =\left(
 \begin{array}{c|c}
  A & B\\
  \hline
  C & D 
 \end{array}\right),
\end{equation}
for complex matrices $A$, $B$, $C$, and $D$ of dimensions $m\times m, n\times m, m\times n$ and $n\times n$ respectively.
From the unitarity of $U$, we have 
\begin{align}
 U\,U^\dagger \equiv \left(\begin{array}{c|c}
  A\,A^\dagger + B\,B^\dagger &A\,C^\dagger + B\,D^\dagger\\\hline
  C\,A^\dagger + D\,B^\dagger& C\,C^\dagger+D\,D^\dagger
 \end{array}\right) &= \mathds{1}_{m+n},\label{Eq:UUDBlocks}\\
U^\dagger U \equiv \left(\begin{array}{c|c}
  A^\dagger A + C^\dagger C&A^\dagger B+ C^\dagger D\\\hline
  B^\dagger A+ D^\dagger C& B^\dagger B+D^\dagger D
 \end{array}\right) &= \mathds{1}_{m+n}.\label{Eq:UDUBlocks}
\end{align}
Considering the blocks on the diagonals of Eqs.~\eqref{Eq:UUDBlocks}, we obtain the matrix equations 
\begin{align}
A\,A^\dagger + B\,B^\dagger &= \mathds{1}_m, \label{Eq:07}\\
C\,C^\dagger+D\,D^\dagger &= \mathds{1}_n.\label{Eq:08}
\end{align}
Equations~\eqref{Eq:07} and~\eqref{Eq:08} imply that
\begin{align}
[A\,A^\dagger, B\,B^\dagger] &= 0,\label{Eq:Commutation1}\\
[C\,C^\dagger, D\,D^\dagger] &= 0,\label{Eq:CDCommute}
\end{align}
i.e., $A\,A^\dagger$ commutes with $B\,B^\dagger$ and $C\,C^\dagger$ commutes with $D\,D^\dagger$. 
Furthermore, $A\,A^\dagger$ and $B\,B^\dagger$ are normal matrices.
Hence, $A\,A^\dagger$ and $B\,B^\dagger$ are diagonalized by the same matrix; or $A$ and $B$ have the same (up to a phase) left-singular vectors, denoted by the unitary matrix $L_m$.
From Eq.~\eqref{Eq:CDCommute}, $C$ and $D$ have the same left-singular vectors, denoted by $L_n'$.

From Eq.~\eqref{Eq:UDUBlocks}, we have
\begin{align}
A^\dagger A + C^\dagger C &= \mathds{1}_m,\label{Eq:09}\\
B^\dagger B+D^\dagger D&= \mathds{1}_n.\label{Eq:10}
\end{align}
Following the same line of reasoning as the one used for obtaining common left-singular vectors, we observe that matrices $A$ and $C$ have the same right-singular vectors, say $R_m$, and $B$ and $D$ have the same right-singular vectors $R'_n$.

 
The left- and right-singular vectors of the matrices $\{A,\,B,\,C,\,D\}$ can be employed to diagonalize these matrices according to
\begin{align}
A &= L_m\Lambda^{A}R^{\dagger}_m,\label{Eq:A}\\
B &= L_m\Lambda^{B}R^{\prime\dagger}_n,\label{Eq:B}\\
C &= L'_n\Lambda^{C}R^{\dagger}_m,\label{Eq:C}\\
D &= L'_n\Lambda^{D}R^{\prime\dagger}_n,\label{Eq:D}
\end{align}
for diagonal complex matrices $\{\Lambda^{A},\Lambda^{B},\Lambda^{C},\Lambda^{D}\}$.
The matrices consisting of the absolute values of the corresponding complex elements of $\{\Lambda^{A},\Lambda^{B},\Lambda^{C},\Lambda^{D}\}$ matrices are denoted by $|\Lambda^A|,\,|\Lambda^B|,\,|\Lambda^C|$ and $|\Lambda^D|$ and comprise the singular values of $A,\,B,\,C$ and $D$ matrices respectively. 
Equations~\eqref{Eq:A} to \eqref{Eq:D} can be combined into a single $(m+n)\times (m+n)$ matrix equation
\begin{align}
\left( \begin{array}{c|c}
  A&B\\
  \hline
  C&D
 \end{array}\right) 
 &= \left(
\begin{array}{c|c}
 L_m & \\
\hline
& L'_n
\end{array}\right)
\left(\begin{array}{c|c}
 \Lambda^A & \Lambda^B \\
\hline
 \Lambda^C & \Lambda^D 
\end{array}\right)
\left(
\begin{array}{c|c}
 R_m^\dagger & \\
\hline
& R^{\prime\dagger}_n
\end{array}\right)\nonumber\\
\implies
U&= \tilde{\mathds{L}}_{m+n}
{\tilde{\Lambda}}_{m+n}
\tilde{\mathds{R}}_{m+n}.
\label{Eq:CombineEqn}
\end{align}
Factorization~\eqref{Eq:CombineEqn} is similar to the CSD because $\tilde{\mathds{L}}_{m+n}$ and $\tilde{\mathds{R}}_{m+n}$ block-diagonal unitary matrices and $\tilde{\Lambda}_{m+n}$ comprises diagonal blocks.
In the remainder of this appendix, we show that $\tilde{\Lambda}_{m+n}$ can be brought into the form of a CS matrix~\eqref{Eq:CSMatrix}, thereby completing the construction of the CSD.

If the matrices $L_m$ ($L_n'$) and $R_m$ ($R_n'$) are calculated from the singular value decomposition of $A$ ($D$), then $\Lambda^A$ ($\Lambda^D$) is a real and non-negative diagonal matrix.
The matrices $L_m$, $L_n'$, $R_m$ and $R_n'$ also diagonalize the matrices $C$ and $D$ resulting in $\Lambda^B$ and $\Lambda^C$.
Unlike $\Lambda^A$ and $\Lambda^D$, which consist of real elements, these matrices $\Lambda^B$ and $\Lambda^C$ are complex matrices in general. 
In other words, the diagonal matrices $\Lambda^B$ and $\Lambda^C$ are of the form
\begin{equation}
\begin{aligned}
 \Lambda^B &= P|\Lambda^B|\\
 \Lambda^C &= -|\Lambda^C|P^\dagger,
 \label{Eq:Lambda-BC}
\end{aligned}
\end{equation}
where $P$ is an $m\times m$ diagonal unitary matrix. 
The phases $P_{jj}$ in Eq.~\eqref{Eq:Lambda-BC} for $C$ are complex conjugates of the phases for $B$ because of the unitarity of $\Lambda$.
 
We can remove the matrix $P$ from $\Lambda^B$ and $\Lambda^C$ by redefining $L_m$ and $R_m$ as 
\begin{align}
 \tilde L_m &= L_m P,\\
 \tilde R_m &= R_m P.
\end{align}
Thus, the Eq.~\eqref{Eq:CombineEqn} can be rewritten as:
\begin{equation}
U = \left(
\begin{array}{c|c}
 L_m P& \\
\hline
& L'_n
\end{array}\right)
\left(\begin{array}{r|r}
 \Lambda^A & |\Lambda^B| \\
\hline
 -|\Lambda^C| & \Lambda^D 
\end{array}\right)
\left(
\begin{array}{c|c}
P^{\dagger}R_m^\dagger & \\
\hline
& R^{\prime\dagger}_n
\end{array}\right)
\end{equation}
or
\begin{equation}
U = \mathds{L}_{m+n} \Lambda_{m+n} \mathds{R}_{m+n}.
\end{equation}
Note that the matrix $\Lambda_{m+n}$ comprises only real elements.
Furthermore, $\Lambda_{m+n}$ is unitary because it is a product $\Lambda_{m+n} = \mathds{L}_{m+n}^{\dagger} U \mathds{R}^{\dagger}_{m+n}$.
Hence, $\lambda_{m+n}$ is an orthogonal matrix.

The orthogonality of the $\Lambda$ implies that any two rows and any two columns of the matrix are orthogonal. 
Therefore, the $2\times 2$ block matrices
\begin{align}
 \Lambda_i &= \begin{pmatrix}
  \Lambda_{i,i} & \Lambda_{i,i+m}\\
  \Lambda_{i+m,i} & \Lambda_{i+m,i+m}
\end{pmatrix} 
\end{align}
is also an orthogonal matrix. 
Any $2\times 2$ orthogonal matrix is of the form
\begin{align}
 \Lambda_i &= \begin{pmatrix}
  \cos\theta_i & \sin\theta_i\\
  -\sin\theta_i & \cos\theta_i
\end{pmatrix} 
\end{align}
for $1\le i \le m$. 

Next we consider the case of $i>m$.
For the matrix $\Lambda^{B}$ all the columns with the index $i > m$ are zero. 
Similarly, for  the matrix $\Lambda^{C}$ all the rows with the index $i>m$ are zero. 
From the unitarity of $\Lambda_{m+n}$, we see that each of the diagonal elements in the last $n-m$ columns and rows of the matrix $\Lambda^{D}$ is unity. 
In summary, the matrix $\Lambda_{m+n}$ is of the form
\begin{align}
\Lambda_{m+n} &= \mathds{S}_{2m} \oplus \mathds{1}_{n-m}
\end{align}
for $\mathds{S}_{2m}$ a CS matrix in the form of Eq.~\eqref{Eq:CSMatrix}.

This completes our procedure for factorizing a given unitary matrix using the CSD.
\textsc{matlab} code for our CSD procedure is available online~\cite{Dhand2015}.

\end{document}